# Ontological Foundations of the Variational Principles and the Path Integral Formalism[1]

Vladislav Terekhovich[2]

**Abstract.** In this paper, I consider the issue of how two mathematical models of modern physics—the variational principles and the quantum path integral formalism—relate to reality. I assume that the observed phenomena are consistent with the calculations because both of these models have some common ontological foundations. According to the hypothesis of the summation of coexisting alternative possibilities, at the quantum level, the system at once moves along all histories that possible in given boundary conditions. The actual history is a mere result of the summation of all these alternative possible histories. This resultant history differs from others with the minimal action and maximal probability. I observe this history as a macroscopic trajectory of the system.

## 1 Introduction

One of the amazing properties of the equations of motion is that real systems follow them with unexplained persistence. The reasons for the effectiveness of these equations are the parts of the overall problem, why the mathematical formalisms are such effective for a description of observable reality.[3] Most of the equations of motion correspond to any physical theory, which offers its own model or interpretation that connects these equations with reality. Besides the equations, the models include a set of notions, axioms, and specific logic. All together these are based, often implicitly, on a certain relation to reality. For example, classical mechanics describes reality as an interaction between the point-objects in absolute and uniform flat space-time, using the concepts of the mass, inertia, and force. The field theory considers reality as the interaction between the fields in different spaces, using the concept of the charge, potential, a vector of field intensity, curvature, and others.

---

[1] The text is translated from the Russian original paper that was published in the book "Mathematics and Reality: Moscow Studies in the Philosophy of Mathematics". MSU, Moscow. 2014 (in Russian).

[2] Institute of Philosophy, Saint-Petersburg University. E-mail: v.terekhovich@gmail.com

[3] Vizgin (2011) compares incomprehensibility of the effectiveness of the equations in analytical mechanics with the inconceivable effectiveness of mathematics in general.

There are two mathematical ways to describe the movement that are not accepted to correlate with any explanatory theory, and even more so, these are not usually associated with reality. These are the extremal or variational (hereinafter, I will be used these names as the equivalent) principles (Polack, 1959, 2010) and the quantum path integral formalism of Richard Feynman (Feynman and Hibbs, 1968). On the one hand, the majority of physicists and mathematicians consider both of these models only as a convenient mathematical formalism, well-chosen to describe phenomena. D'Alembert, Lagrange, Jacobi, Einstein, Mach, Prigogine and others objected to any philosophical interpretation of the variational principles. Today, this tradition continues (Lipkin, 2010; Lemons, 1997; Stöltzner, 2009; Yourgrau and Mandelstam, 2000). The attitudes towards a realistic interpretation of the path integral formalism and virtual particles in Feynman diagrams are ambiguous (Gell-Mann and Hartle, 2012; Ogborn and Taylor, 2005; Ord and Gualtieri, 2002; Sharlow, 2007; Taylor, 2003; Valente, 2011). On the other hand, modern philosophers of mathematics are almost not interested in applied mathematics (Steiner, 2005). They are usually limited to pure mathematics while marveling at the miracle of its applicability in the world (Lolly, 2012).

There are several ways to explain why the predictions of the equations are consistent with the observed movement. These can be combined into four groups, arranged from the extremal anti-realism to the extremal realism. (1) In the phenomena themselves, there are no order and logic. People observe the phenomena and create some models to describe the observations, using the common logic and common language. It is not surprisingly that the results of the different people are the same. (2) In real phenomena, there are some laws, but we hardly could know these since any perception is limited by our psycho-physiological features. Therefore, when we create any mathematical model, we adjust them to our perception. (3) The mathematical models reflect the real relationship between different kinds of objects and phenomena; it means that the models describe the phenomenological laws of specific areas of nature. However, the models say nothing about the essence behind the phenomena. (4) The mathematical models are ontological and reflect the real relationships between essences of phenomena.

From a practical point of view, the position of the anti-realist is more convenient. For the anti-realist, any theory is only a temporary model that is suitable to describe a certain kind of the phenomena. The realist faces more difficult task; he has to make a choice between the equations of different models, which of these most sufficiently describe the reality. If the variational principles and the path integral formalism are not directly related to the reality, why we so confident in the models of the reality of classical mechanics or the field theory. Perhaps, there is another option: the models are merely the consequences of the equations of some other, more fundamental theory of real processes. For example, the theory of the mathematical universe states that any physical reality is completely determined by mathematical structures, and, therefore, there can be any mathematically consistent reality (Tegmark, 2008).

The extremal principles and the calculus of variations are now widely distributed not only in optics and mechanics, but also in all versions of the field theory (non-relativistic and relativistic, classical and quantum), in equilibrium and non-equilibrium thermodynamics, in the information theory, in biology, and in other sciences (Terekhovich, 2013). The path integral formalism and "Feynman diagrams" lie in the base of modern quantum field theory (Zinn-Justin, 2010). It is known the formal mathematical relationship between some variational principles and the path integral formalism (Hanc, Tuleja and Hancova, 2003; Ogborn and Taylor, 2005; Taylor, 2003). The attempts to accept the principle of least action as a general principle of nature were undertaken by Leibniz, Maupertuis, Euler, Helmholtz, Max, and Planck (Polack, 1959). Some Russian philosophers have tried to link the extremal principles with the laws of dialectics (Myakishev, 1973; Asseev, 1977; Razumovsky, 1975; Tsekhmistro, 1992). Some modern authors consider a realistic approach to the quantum path integral (Gell-Mann and Hartle, 2012; Valente, 2011; Wharton at al., 2011; Ord and Gualtieri, 2012; Sharlow, 2007). In this paper, we make an attempt to combine two models of the mathematical description– the variational principles and the path integral formalism in their realistic interpretation. Perhaps, if we understand how mathematics works in these specific applications, we are closer to the solution of common issues in pure mathematics.

The paper starts with an overview of the features of the calculus of variations and the path integral formalism (Section 2). After listing some of the philosophical issues associated with these models, we consider the search direction of ontological grounds of the models (Section 3). In Section 4, we consider the hypothesis of the summation of alternative coexisting alternative possibilities, where the ontological assumptions are combined with the mathematical description of motion. The derived results are used for explanation of the relationship between the equations of motion and reality (Section 5).

## 2 Features of variational principles and path integral formalism

If we get acquainted with the history of the variational principles (Asseev, 1977; Polack, 2010), as well as with the story how Richard Feynman came to the formulation of the path integral formalism (Feynman, Leighton, Sands, 2004, p. 96), we can note some interesting features of these two very different models that describe the motion. Let list only the main of the features.

1. The variation or extremal principles lie at the heart of modern natural science. These are suitable for the description of linear or nonlinear processes in the closed or open systems of varying complexity, from elementary particles to social systems. The principles are also applied to the geometry and topology of the different dimensions. In additional, in non-equilibrium thermodynamics and information theory, these principles involve the concept of probability.

2. The integral variational principles can be reduced to a single scheme: the actual process (or path) differs from all alternative possible processes, consistent with the given constraints, that its own functional (for example, *the action*), which describes the system, is stationary and takes an extremal value. In most cases, it is a local minimum. However, it also may be a maximum. The functional is defined as an integral of a certain expression (called the *Lagrangian* or *Lagrangian density*), and can be calculated over the path, time, n-dimensional volume, or four-dimensional space-time.[4] In the calculus of variations, the extremum that corresponds to the actual movement or state is sought by the operation of varying or examination of all conceivable movements or states, which are not actualized in a reality. The difference between the actual and any possible value of the functional in the first order of approximation must be zero. The differential equations of motion and the equation of state systems could be derived from the variational principles.

3. The variational principles describe some stationary processes (path or state) in n-dimensional configuration space, along which the system tends to follow in any given conditions. One of the cases of stationarity is the constancy of the speed the system functional changes. A special case of the stationary process is an equilibrium process; a special case of the equilibrium process is an equilibrium state.

4. Most of the variational principles are related to each other through the analogy of mechanical, optical and wave phenomena. The analogy is not only used on the classical level, but also on the relativistic and quantum ones. Most of the variational principles and the path integral formalism use the concept of *the action* that has a dimension of energy multiplied by time.

5. The path integral formalism is based on Feynman's assumption that quantum particles at once move from an initial state to a final state from along all possible paths. The observed or actual path is the result of the summation of all possible paths. The summation occurs thanks to the interference or adding of the probability amplitude phases of each possible path. Thus, the resulting path has a maximum probability. When a scale of the quantum systems increases, the path integral formalism formally transfers to one of the basic physical variational principles—the classic principle of least action.

6. The relationship between the principles of symmetry and principles of conservation based on the variational principles (Noether's theorem), and even the law of conservation of energy is a consequence of the equation variation and the invariance of the action.

7. The variational principles, as well as the path integral formalism use, in the strict mathematical sense, two fundamental philosophical concepts—*potential possibility* and *actual reality*.

8. In the variational principles, the descriptions of real physical processes simultaneously use as well the initial as final states of the system. In other words, unlike all other laws, these models do not give any preference to neither efficient nor final causes.

---

[4] To describe the actual state of the system, in the differential variational principles, instead of the integration it is used the summation that set equal to zero.

# 3 Research program in the quest of ontological foundations

The common properties of the variational principles and the path integral formalism do not explain their effectiveness yet. On the contrary, many new questions arise. The major part of these questions lie in the field of philosophy, and even their own formulations are difficult. Let us list some examples of such questions.

What is the physical and philosophical status of the potential or virtual movements (trajectories, states) in the variational principles? What is the degree of their reality? The similar questions concern the virtual paths in the path integral formalism. What do the classical and quantum actions mean in the physical and philosophical sense? Why does the action strive to extremal values? Why, in certain cases, the action is minimal, but, in other cases, it is maximum? How is the classical action related to the quantum action and the tensor of space curvature? What is the philosophical and physical content of complex variables in the equations of motion? Why are properties of symmetry and conservation linked with the classical action and the path integral formalism? How do the extremal principles relate to the efficient and final causes? Why are the extremal principles equally effective in the description of probabilistic and deterministic processes? How does the transition occur from the probability amplitude at the quantum level to the probability and the uniqueness of the classical level?

A simple review of the properties of the variational principles and the path integral formalism suggests that their mutual relationship is not a coincidence. Moreover, we can suppose that together they can be a good candidate for the investigation of possible ontological foundations of the equations of motion. However, we need more than only intuition. To build a consistent hypothesis, it is necessary to investigate the fields of philosophy, physics, and mathematics that are very distant from each other. The program of such investigation, as a minimum, should include:

- the philosophical concept that a possible or potential mode of existence transforms itself into an actual mode of existence;

- the idea that a probabilistic description of all processes in nature is the most fundamental, and the deterministic or statistical reasons are mere special cases of probabilistic causality;

- philosophical perspectives on the equality between efficient and final reasons as two complementary aspects of causation;

- the concept that probability is a measure of the transition from possibility into actuality;

- the old tradition to consider an internal activity of any physical system as ontological property and a probability as a consequence of self-motion of matter;

- some modern interpretation of quantum mechanics that, in different ways, use the notions of possible and actual existence: Copenhagen, holistic, consistent histories, modal, many-worlds, existential, and others;

- mathematical features of the path integral formalism to calculate the trajectories and states of the quantum particles and fields;

- the correlation between the quantum probability amplitudes in the path integral formalism, the classical theory of probability, and theory of information;

- the correlation between the usage of complex values and natural logarithms in the path integral formalism and the equations of motion of classical systems.

We believe that this research program will lead to a hypothesis that there is a common ontological basis of different kinds of the motion equations, including the equations of the variational principles and the path integral formalism. Most likely, there will be several hypotheses, and each of them has to give its version of consistent answers to the questions above.

## 4 Summation of coexisting alternative possibilities

In one of the attempts to follow the proposed research program (Terekhovich, 2013) we have formulated *a hypothesis of the summation of coexisting alternative possibilities*. An ontological layer of this hypothesis is based on three assumptions.

**Internal activity.** Let us assume that each of the real systems has a certain degree of internal activity. Activity is expressed in a tendency of the system to implement the maximal number of its possibilities to change and maintain the current motion or state. This implementation is due to the interactions of every system with other systems through the modification of existing connections and the creation of new mutual ones.[5]

**Two modes of existence**. Let us assume that each of the real systems exists in two modes of existence. In a potential mode of existence, the real system is in all possible alternative motions at once and interacts with other systems in all possible ways. In the actual mode of existence, the system is in one of the possible alternative motions. Every possible motion is described by the characteristics of n-dimensional spaces with various mathematically possible topologies. Every actual motion is described by the characteristics of smooth four-dimensional space-time. The set of the possible alternative motions constitutes a possible reality; a set of the actual motions constitutes an actual reality. Two realities exist "in parallel" transforming into each other.

**The summation of alternative possibilities**. Let us assume that a package of the consistent possible alternative motions of the system is continuously added together. One of the possible motions that combines the most number of alternative possibilities becomes the actual motion. Other alternative possible motions of the system continue to exist in the possible mode. Changes or disappearance of at least one possible motion can change the result of the summation within the whole package. Thus the actual motion can change. The interaction of the actual systems does not happen in the actual mode; it happens in the potential mode of existence. Each of the interactions changes the packages of the possible alternative motions of the system. This automatically leads to changes in the actual motion of the interacting system.[6]

---

[5] This assumption has a rich philosophical tradition and can be used to analyze the general principles of the existence of physical, biological, social and psychological structures (see. Terekhovich, 2012).

A physical layer of the proposed hypothesis translates these ontological assumptions into the language of physical concepts.[7] Let us start with the quantum systems, which are quite accurately described by the path integral formalism. If the assumption of two modes of existence is true, the quantum system actually moves at once along all alternative histories (trajectories) that are possible in given boundary conditions. These histories are in coherent superposition since these have consistent phases of probability amplitudes. If the assumption about the summation of alternative possibilities is true, the package of all possible alternative quantum histories adds together thanks to the interference in n-dimensional configuration space. It means that all phases of such histories are added. The phase of probability amplitudes is a complex quantity that is proportional to the action. To get the total probability of the actual history, it is necessary to be squared the modulus of the sum of the probability amplitudes within the possible histories package. A contribution of every possible alternative history is proportional to its own phase. Thus the resulting history differs from others that it has the minimal action and maximal probability. This resultant history of the quantum system is that we observe using macroscopic devices.

To pass from quantum systems to classical ones, we have to assume that there are no compelling ontological boundaries between the micro and macro objects. Of course, there are some important differences, but these are not absolute. At a fundamental level, all systems exist under the same principles. The observed differences are rooted in the peculiarities of our perception and description of the existence of various systems with different sizes. At the classical level, we can describe the systems only in terms of the actual existence, and we can do this with a certain degree of approximation. At the quantum level, we cannot already ignore the possible alternative states or histories and their influence on the actual existence.

---

[6] A similar hypothesis was formulated by Leibniz. He stated that substances are beings capable of action; it means that they are endowed with primitive active and passive powers (Leibniz, 1989, p. 159, 207). Then, he introduced a theory of the striving possibles (Leibniz, 1951, p. 347-349), where he showed a distinction between essence (the nature of a thing) and existence. He postulated that the principle of governing essences is that of possibility or non-contradiction. Every essence (possible thing) tends of itself towards existence, but the one that will actually exist is that which has the greatest perfection or degree of essence or the greatest number of possibles at the same time. Leibniz (1982, P. 235-284) postulated the principle of the greatest amount of existence that explains why if you want to go from one point to another, then you select the easiest and shortest way.

[7] A review of the philosophical arguments in favor of the ontological many-modes model of reality for quantum phenomena is given in (Sevalnikov, 2009).

Let us combine three ontological assumptions with the description of the quantum objects. Consider how the description of reality can change when the size of the system becomes significantly greater than the length of the wave function. The package of the possible alternative histories that make a substantial contribution to the actual history shrinks into a narrow beam. The fluctuating quantum n-dimensional space collapses into a smooth four-dimensional curved space-time. The quantum possible alternative histories are transformed into the virtual motions (trajectories, states) of the classical systems. The quantum action in the limit gradually reduces to the classical action; the path integral formalism reduces to one of the variational principles. It follows that the package of quantum alternative histories reduces to the observed classical trajectory along a geodesic path with the minimal action (or other functional's extremum). Thus, we can say that the maximal quantum probability is obtained when the phases of probability amplitudes of the possible alternative histories are added together. The maximal probability manifests in the classical and relativistic limits as a minimum or a maximum of one of the system's characteristics that is reflected in one of the variational principles.

Summing up, we assume that all variational principles basically have a probabilistic nature and the common ontological source at the quantum level. Given the mutual relationships between the various variational principles, let us call such conclusion the probabilistic interpretation of the principle of least action. This hypothesis, despite its strangeness, can explain a surprising prevalence of the variational principles in science.[8]

## 5 How equations are related to reality

Let us check whether the hypothesis of the summation of coexisting alternative possibilities offers some new answers to the issue of predictive power of the equations of motion. According to this hypothesis, the minimum of the action of the actual motion loses its mystery and presents a simple consequence of the mechanism of interference at the quantum level. The minimum in the principles of least action can be considered as special cases of maximal probability. Thus, this principle can be explained through its probabilistic interpretation.

It is known that every variational principle corresponds to the differential equation of motion and vice versa. This equivalence does not still explain what is primary or secondary—the variational principles or differential equations. However, if we assume that the variational principles are based on the probabilistic interpretation of the principle of least action, then these have a common foundation at the quantum level. Thus, the differential equations can be considered as one of the mathematical forms of the variational principles.

---

[8] In 1920, Eddington expressed a similar idea. He believed that the principle of least action is the principle of the most probability. The law of nature is that the state of the world, which is implemented in reality, is the most probable state, and physical reality is a synthesis of all possible physical aspects of nature (Eddington, 2003).

In the variational principles, the notions of the possible alternative trajectories or possible motions are habitual. They are also called *virtual* or *imaginable*. In light of the hypothesis of the summation of coexisting alternative possibilities, these notions (rather strange for physics) are not metaphors, but the words with certain metaphysical content. Possible or virtual alternative motions occur in reality but in the potential mode of existence only. From the viewpoint of the actual existence, this statement looks absurd since one actual system cannot simultaneously be in the different places of the four-dimensional space-time. In the potential mode of existence, there is no contradiction if the four-dimensional space-time is regarded as a consequence of the interaction of the actual systems but not like something given in advance. The virtual motions are not a figment of our consciousness; these are reflections of the alternative possibilities. In other words, the variations in the variational principles do not take place in the mind of a mathematician (there these occur too). These take place in the potential mode of existence. Mathematicians use them to calculate the actual motion, and then they surprise that their calculations coincide with observations.

We based on a metaphysical idea that space is not an arena, where actual systems implement their actual interactions; rather this is the form and the result of these interactions[9]. We assume that the geometric properties of space are defined by features of a mechanism, under which the potential mode of existence passes into the actual one. In Gauss's principle of the least constraint, the value of the constraint is equivalent to a geodesic curvature of a point's trajectory in three-dimensional Euclidean space. In a geometric analogue of Gauss's principle—a Hertz's principle—the point tends to minimize the curvature of its own trajectory. The principle of least action is successfully used in a modern superstring theory for spaces with many dimensions. This suggests that the geometrical properties of space (Euclidean, Riemannian, or Finslerian) might be associated with the mathematical features of the mechanism of interference between the systems' possible alternative motions at the quantum level.[10]

One of the metaphysical issues is related to the reality of a mathematical object called a probability. There is a psychological view of the probability as the most plausible or expected outcome of the affair. There are also several quite scientific interpretations of the probability, for example, statistical, information, and quantum. The first one considers the probability as an average frequency of a set of events that already occurred and are observed. The second one connects the probability with a degree of uncertainty or a degree of ordering of complex systems. The third one considers the probability as a measure of propensity some quantum event to be observed in actuality. In our hypothesis of the summation of coexisting alternative possibilities, the probability is considered as the measure of the implementation of the specific possibility in the actual mode of existence. Every possible alternative history is continuously involved in the formation of the actual history, so the contradiction between the quantum and statistical interpretation of probability disappears. The fact is that it does not matter whether we study the measure of implementation from the point of view before the implementation or after this—from the present into the future or from the present into the past. The measure of implementation of the specific possibility always is in the present in each possible and each actual history.

---

[9] Leibniz argued the conception that space is derivative from interaction of objects (Leibniz, 1982, p. 325).

[10] Eddington proposed to consider the action as a counterpart to the curvature of space (Eddington, 2003).

Another consequence of the hypothesis of the summation of coexisting alternative possibilities is associated with the ontological interpretation of complex numbers. Probability amplitude of each alternative quantum history contains a complex phase. It means that the phase has both real and imaginary parts. After interference and the summation of these phases, the package of alternative histories turns into the actual history, so the imaginary parts of the phases disappear. This suggests that the meaning of the imaginary part of the quantum phase is connected with the potential mode of existence; and a mathematical operation of squaring the modulus of the probability amplitude describes the transition from the many possible alternative histories into one actual history.

Assume that the quantum level is fundamental for any kinds of events, then various oscillations and waves that we observe in the actual world have to be connected with the wave functions and probability amplitudes of the quantum objects. Feynman represented the probability amplitude as a vector that rotates in an abstract space and the quantum phases as angles of this rotation. In the classical limit, this representation is similar to Huygens' method of calculating of a wavefront through the summation of microwaves. At the same time, this is similar to Fermat's principle for beams of light that propagate along the geodesic lines. Take also into account the idea that Hamilton's principle of classical mechanics is analogous to Fermat's principle because the material point moves along the line orthogonal to the front of the phase wave in configuration space.[11] Perhaps this explains why all the oscillations and waves can be described by the same mathematical means. However, to accept this explanation we have to sacrifice some of the settings of common sense. We have to assume that Feynman path integral, Huygens', Fermat's, and Hamilton principles are not mere products of our mind, but also the models that adequately describe one of the aspects of reality. As a physicist Michio Kaku said, our representation of the physical universe based on common sense is merely the most probable state from an infinite number of possible states; we coexist with all possible alternative states, some of them could move us into the age of dinosaurs, to a nearby supernova, or to the bound of the universe (Kaku, 2007).

---

[11] De Broglie and Schrödinger used this analogy to create a wave quantum mechanics (Polak, 1959, p. 691, 861).

The next step is the conclusion that the universal laws of the oscillations and waves are determined by the same rules of interference between possible alternative histories of various kinds of objects at the quantum level. The unity of these rules is also reflected in the general principles of symmetry and conservation. It is due to these general rules of the interference of the coexisting possibilities; our classical space is homogeneous, isotropic, and three-dimensional. Due to these general rules, the objects of the same kind obey the same laws. If this is true, the fact that the actual world seems so orderly, simple and beautiful is not surprising. It cannot be otherwise because from a mathematical viewpoint the result of interference between the alternative possibilities cannot be different, at least in our universe. Theoretically, one can imagine that in other universes there are other rules of interference between the alternative possibilities. Then there must be other principles of symmetry, conservation, and extremality. Then there the actual systems would otherwise interact with each other, and space would have different properties.

Finally, the hypothesis of the summation of coexisting alternative possibilities may be useful in a discussion about the status of actual infinity in mathematics. It is known that Cantor tried to reduce physics to the theory of point sets. He referred to the monads of Leibniz as the prime elements of nature, and all matter emerges from the union of these elements (Cantor, 1985, p. 168). Let the package of an infinite set of the alternative possible histories is defined as a potential infinity of the complex phases of the probability amplitudes, which describe every possible history. Then the result of interference between the possible histories (summation of the phases)—the actual history—can be considered from two points of view: (a) as the finite set of actual states in terms of four-dimensional space-time; (b) as the infinite set of the possible alternative histories and the possible alternative states in given boundary conditions. Thus, every aspect of the actual mode of reality contains the infinite set of the aspects of the potential mode of reality. Infinity becomes a necessary link between potential and actual modes of reality.

## 6 Conclusion

The subject of this paper is the issue of how two mathematical models that widely used in modern physics—the variational principles and the quantum path integral formalism—relate to reality. We suggest that the observed phenomena are consistent with the calculation because both of these models have some common ontological foundations. According to the hypothesis of summation of coexisting alternative possibilities, the system at once moves along all histories that possible in given boundary conditions, and the actual history (or actual state) is the sum of all alternative possible histories (or possible alternative states). The actual history (actual state) has the maximum probability.

This hypothesis raises a number of objections that can be divided into two groups. The philosophers generally oppose the transfer of any mathematical and physical models to reality. However, as history shows, most of the important theories of reality use, in varying degrees, the models of geometry and the natural sciences. In the 19th century, the main sources of concepts of reality were Darwin's theory and theory of electromagnetism. In the 20th century, it is hardly possible to find some ontological concept that is not directly or indirectly influenced by three physical models, which were created solely for the convenience of mathematical calculations. We mean the general and special relativity, the model of the expanding universe, and quantum mechanics with the principles of uncertainty and complementarity, the EPR paradox, and Schrödinger's cat.

Another objection comes from the physicists. They are not against the transfer of the physical models to reality, but they agree only with models where there are objects with a clear physical meaning. It is commonly argued that the variational principles and the path integral formalism do not have any physical content; these are merely convenient metaphors. However, the presence or absence of physical content is the arguable argument because the content itself is our model. For instance, how to decide, which model possesses more physical content: the force of gravity acting at a distance; curved space-time; the electron as a particle or as a cloud of probability. Another objection is that variational principles and the path integral formalism have many limitations in practical applications. One of the answers is that the purpose of our hypothesis is not a representation of the Feynman path integral as the ontological theory. On the contrary, the purpose is to investigate the ontological bases, which do explain the path integral formalism and also outline possible solutions to other issues. The point is not whether legally or not we correlate the successful physical and mathematical models with ontological constructs of reality. The point is to test and compare the various ways. One of the most famous examples of this testing and comparison are the various interpretations of quantum mechanics and quantum cosmology.

The research program is dedicated to the issue of how the equations of the variational principles and the path integral formalism are connected with reality. This program can give not only one result. They might create many hypotheses—more convenient than the hypothesis of summation of coexisting alternative possibilities. Only the competition of old, contemporary, and future hypotheses can develop our understanding of reality. If we assume that mathematics is not limited by our brain, but in some degree reflects the real processes, we should use the mathematical operations to refine and systematize the ontological structures. After this, the mathematicians can hope that ordered ontological ideas will help them to explain the meaning of familiar mathematical objects.